\documentclass[12pt,floatfix,showkeys,superscriptaddress,aps,prd,preprint]{revtex4}
\usepackage[latin1]{inputenc}
\usepackage[T1]{fontenc}
\usepackage{lmodern}
\usepackage{amsmath}
\usepackage{amssymb}
\usepackage{graphicx}
\usepackage{esint}
\usepackage{longtable}
\usepackage{dcolumn}
\usepackage[english]{babel}
\usepackage{csquotes}
\usepackage{color}
\usepackage{slashed}
\usepackage{simplewick}
\usepackage{tikz-feynman}
\usepackage[]{tikz-feynman}
\usepackage{amsmath,latexsym}
\usepackage{color}

\usepackage{hyperref}
\hypersetup{
    colorlinks,
    citecolor=blue,
    filecolor=green,
    linkcolor=purple,
    urlcolor=red,
}

\usepackage{slashed}

\usepackage{hyperref}
\hypersetup{colorlinks,breaklinks,
			citecolor=[rgb]{0,0.0,1.0},
            urlcolor=[rgb]{0.0,0.0,1.0},
            linkcolor=[rgb]{0,0.5,0.9}}

\def\be{\begin{equation}}
\def\ee{\end{equation}}
\def\bea{\begin{eqnarray}}
\def\eea{\end{eqnarray}}

\begin{document}


\title{5D Elko spinor field non-minimally coupled to nonmetricity in $f(Q)$ gravity}

\author{F. M. Belchior}
\email{belchior@fisica.ufc.br}
\affiliation{Universidade Federal do Cear\'a (UFC), Departamento de F\'isica,\\ Campus do Pici, Fortaleza - CE, C.P. 6030, 60455-760 - Brazil.}
\author{A. R. P. Moreira}
\email{allan.moreira@fisica.ufc.br}
\affiliation{Universidade Federal do Cear\'a (UFC), Departamento de F\'isica,\\ Campus do Pici, Fortaleza - CE, C.P. 6030, 60455-760 - Brazil.}
\author{R. V. Maluf}
\email{r.v.maluf@fisica.ufc.br}
\affiliation{Universidade Federal do Cear\'a (UFC), Departamento de F\'isica,\\ Campus do Pici, Fortaleza - CE, C.P. 6030, 60455-760 - Brazil.}
\affiliation{Departamento de F\'{i}sica Te\'{o}rica and IFIC, Centro Mixto Universidad de Valencia - CSIC. Universidad
de Valencia, Burjassot-46100, Valencia, Spain.}

\author{C. A. S. Almeida}
\email{carlos@fisica.ufc.br}
\affiliation{Universidade Federal do Cear\'a (UFC), Departamento de F\'isica,\\ Campus do Pici, Fortaleza - CE, C.P. 6030, 60455-760 - Brazil.}



\begin{abstract}
This paper aims to investigate the localization of the five-dimensional spinor field known as Elko (dual-helicity eigenspinors of the charge conjugation operator) by employing a Yukawa-like geometrical coupling in which the Elko field is non-minimally coupled to nonmetricity scalar $Q$. We adopt the braneworld scenarios in which the first-order formalism with sine-Gordon and linear superpotentials is employed to obtain the warp factors. A linear function supports the zero-mode trapping within the geometric coupling, leading to the same effective potential as the scalar field. Moreover, an exotic term must be added to obtain real-valued massive modes. Such modes are investigated through the Schr\"{o}dinger-like approach. 
\end{abstract}

\maketitle

\section{Introduction}
There is a consensus that general relativity (GR) needs to be modified to explain physics at both quantum and galactic scales. In this sense, modified gravity theories have gained considerable interest in the literature over the past years. Such theories can be constructed by adding geometrical invariant, such as a generalized function of curvature scalar $R$ \cite{Capozziello:2005ku,Bazeia:2014poa,Gu:2014ssa}, energy-momentum tensor $\mathcal{T}$ \cite{Myrzakulov:2012qp,Deb:2017rhc,Moraes:2016akv}, Gauss-Bonnet term $G$ \cite{Li:2007jm,Elizalde:2010jx,SantosDaCosta:2018bbw},  among others, into Einstein-Hilbert action. In particular, $f(R)$ gravity has been investigated as a possible mechanism to explain the recently observed cosmic acceleration and could describe dark matter
\cite{DeFelice:2010aj, darkmatter, cosmologicalconstant, Bisabr:2010xy}. 

It is still possible to construct a modified gravity wherein the gravitational dynamics is encoded by torsion scalar $T$ rather than curvature. This theory is known as teleparallel equivalent of general relativity (TEGR) \cite{deAndrade1999, Aldrovandi, Bahamonde:2021gfp}. Like the GR, it can be directly extended by assuming a general function of torsion scalar, resulting in $f(T)$ gravity, which has been intensively worked in the literature \cite{Ferraro2011us, Tamanini2012, Yang2017, ftenergyconditions, Bahamonde2015,Tan:2020sys,Wei:2011yr}. Another curvature-free modified gravity is the symmetric teleparallel equivalent of general relativity (STEGR) \cite{Nester:1998mp}, whose gravitational action depends on nonmetricity scalar $Q$. Such a gravity and its direct extension $f(Q)$ gravity have gained highlight in several contexts, such as wormhole, black hole, dark matter, and cosmology \cite{Ayuso:2020dcu,BeltranJimenez:2019esp,BeltranJimenez:2017tkd,BeltranJimenez:2019tme,Bajardi:2020fxh,Capozziello:2022tvvi,Capozziello:2022wgl}. 

On the other hand, several kinds of topological defects can build a thick brane model, where our Universe is a membrane embedded in a warped five-dimensional spacetime \cite{rs,rs2,Gremm1999,Afonso2006,Janssen2007,Menezes,Bazeia:2003aw,Bazeia:2004dh, Bazeia:2021jok, Bazeia:2021bwg, Moreira:2021xfe, Moreira:2021uod, Belchior}. This work aims to investigate the Elko field localization in braneworlds models assuming a $f(Q)$ gravity \cite{Fu:2021rgu, Silva:2022pfd}. The Elko field is a spinor of half spin with mass dimension one (four-dimensional spacetime), which is the eigenspinor of the charge conjugation \cite{daRocha:2011yr, Ahluwalia:2008xi, Ahluwalia:2022ttu, Ahluwalia:2009rh, Ahluwalia:2010zn, Fabbri:2010ws}. This field has been employed as the first fermionic field to describe dark matter, besides being regarded as a dark field since it does not interact with the electromagnetic field. In this sense, only the graviton and Higgs field can interact with the Elko field. In Refs. \cite{Pereira:2016muy, Maluf:2019ujv} the Casimir energy was analyzed for the Elko field, confirming its fermionic nature theoretically with a repulsive Casimir force. Recently, it has been proposed some attempts to detect the Elko field at the Large Hadron Collider (LHC) \cite{Dias:2010aa, Alves:2014kta}.

The confinement of the Elko spinor in Randall-Sundrum scenarios was addressed for the first time in \cite{Liu:2011nb}, showing that the Elko field, like other fields, requires a suitable coupling to be trapped on a five-dimensional braneworld. In this work, Yukawa-like coupling between the Elko spinor and the background scalar field was proposed. Other work were proposed by considering Yukawa-like \cite{MoazzenSorkhi:2020fqp,Zhou:2020ucc} and dilaton-like coupling \cite{Zhou:2017bbj,Sorkhi:2018jhy}. In addition, a geometric coupling with the Ricci scalar was considered in Ref. \cite{Jardim:2014xla}. However, there is a common issue in these works: a complex value potential, which makes studying massive and resonant modes difficult. In the context of string-like brane, such an issue can be overcome by adding an exotic term, as shown in Ref. \cite{Dantas:2015mfi}. Our investigation will consider a geometrical coupling between the Elko field and the nonmetricity scalar through a Yukawa-like interaction in 5D thick braneworld. This coupling leads to normalizable zero-mode besides providing real-valued massive modes \cite{Silva:2022pfd}.

This paper is organized as follows: In section (\ref{s2}), the main concepts of $f(Q)$ gravity are briefly discussed. Next, the five-dimensional thick brane is studied in section (\ref{s3}), and the first-order formalism is introduced to obtain analytical solutions. In section (\ref{s4}), a Yukawa-like coupling is proposed to study the localization of Elko field zero-mode as well as its resonates massive modes. The conclusion of this present work and future perspectives are discussed in section (\ref{s5}).

\section{$f(Q)$ gravity}\label{s2}

A brief review of symmetric teleparallel gravity as well as the equations of motion for $f(Q)$ gravity, is presented in this section. Let us initially distinguish essential concepts of GR and STEGR. An important feature of Riemannian geometry is the metricity condition given by
\begin{equation}\label{mc}
\nabla_M g_{NP}=0,    
\end{equation}
where $g_{NP}$ is the metric and $\nabla_M$ is the covariant derivative with the Levi-Civita $\Gamma^P\ _{MN}$ as affine connection.
Such condition is shared by GR and its direct modification as $f(R)$ gravity. We denote the bulk coordinate indices by capital Latin index $M=0,\ldots,D-1$.

On the other hand, since STEGR is based on nonriemannian geometry, the relation (\ref{mc}) is no longer satisfied, leading to nonvanishing nonmetricity tensor \cite{Nester:1998mp}
\begin{equation}
Q_{MNP}=\nabla_M g_{NP},    
\end{equation}
which has the following independent traces
\begin{eqnarray}
Q_M=g^{NP}Q_{MNP},\\
\widetilde{Q}_M=g^{NP}Q_{NMP}.
\end{eqnarray}
The more general connection $\widetilde{\Gamma}^P\ _{MN}$ for STEGR is then defined as
\begin{equation}
\widetilde{\Gamma}^P\ _{MN}=\Gamma^P\ _{MN}+L^P\ _{MN},    
\end{equation}
where $L^P\ _{MN}$ is defined as the distortion tensor, which is written in nonmetricity tensor terms as \cite{Nester:1998mp}
\begin{equation}
L^P\ _{MN}=\frac{1}{2}g^{PQ}(Q_{PMN}-Q_{MPN}-Q_{NPM}).    
\end{equation}
At this point, it is convenient to introduce a more general tensor that contains the nonmetricity, its independent traces, and distortion tensor. Such tensor is known as nonmetricity conjugate given by
\begin{equation}
P^P\ _{MN}=-\frac{1}{2}L^P\ _{MN}+\frac{1}{4}(Q^P-\widetilde{Q}^P)g_{MN}-\frac{1}{8}(\delta^P_M Q_N+\delta^P_N Q_M).    
\end{equation}
Besides, its contraction with nonmetricity tensor provides the nonmetricity scalar $Q=Q_{PMN}P^{PMN}$. The Ricci scalar is written as $R=Q+B$, where $B$ is a boundary term. Such result shows that STEGR is equivalent to GR since the boundary term vanishes when integrated in the action. 

The $f(Q)$ gravity represents a direct extension of STEGR. The five-dimensional gravitational action for this gravity is assumed as \cite{Fu:2021rgu, Silva:2022pfd}
\begin{equation}\label{a1}
S=\int d^5x \sqrt{-g}\left(\frac{1}{2}f(Q)+\mathcal{L}_m\right),
\end{equation}
where $\mathcal{L}_m$ represents the matter Lagrangian to be defined in the next section. The variation of action (\ref{a1}) with respect to the metric gives the following equation
\begin{equation}
\frac{2}{\sqrt{-g}}\nabla_K(\sqrt{-g}f_QP^K\ _{MN}) -\frac{1}{2}g_{MN}f+f_Q(P_{MKL}Q_N\ ^{KL}-2Q^ L\ _{KM}P^K\ _{NL})=\mathcal{T}_{MN},  
\end{equation}
where $\mathcal{T}_{MN}$ is the energy-momentum tensor. Furthermore, if we vary the action (\ref{a1}) with respect to the connection, one gets
\begin{equation}
\nabla_M\nabla_N(\sqrt{-g}f_Q P_K\ ^{MN})=0.   
\end{equation}
Here, we set $f\equiv f(Q)$ and $f_{Q}\equiv \partial f(Q)/\partial Q$ for simplicity.

\section{Thick brane scenarios and First-order formalism}\label{s3}

This section is dedicated to constructing the braneworld scenarios in $f(Q)$ gravity. Let us consider a single scalar field as a matter source that generates the thick brane. Thus, the matter Lagrangian is given by
\begin{equation}
\mathcal{L}_m=-\frac{1}{2}\partial_M\phi\partial^M\phi -V(\phi).   
\end{equation}
The energy-momentum tensor associated to this Lagrangian reads
\begin{equation}
T_{MN}=\partial_M\phi\partial_N\phi+g_{MN}\mathcal{L}_m.    
\end{equation}

Let us now use the ansatz for a generic Randall-Sundrum-like metric as \cite{rs,rs2}
\begin{equation}\label{metric}
ds^2= e^{2A}\eta_{\mu\nu}dx^{\mu}dx^{\nu}+dy^2,    
\end{equation}
where $\eta_{\mu\nu}$ is the Minkowski metric, $e^{2A}$ is the warp factor, and $y$ represent the extra dimension. Here, the coincident gauge, i.e., $\widetilde{\Gamma}^P\ _{MN}=0$ is also considered. Then, for this metric the nonmetricity scalar is $Q=12A^{\prime 2}$, while the scalar field and gravitational equations read
\begin{eqnarray}
\phi^{\prime\prime}+4A^{\prime}\phi^{\prime}&=&V_{\phi},\\
\label{e.q} A^{\prime}f_Q^{\prime}+f_Q A^{\prime\prime}&=&-\frac{\phi^{\prime 2}}{3},\\
\label{e.q2} 12f_QA^{\prime 2}-\frac{f}{2}&=&\frac{\phi^{\prime 2}}{2}-V.
\end{eqnarray}

In the context of braneworlds and topological structures, analytical solutions for equations of motion can be obtained by employing the so-called first-order formalism \cite{Gremm1999,Afonso2006,Janssen2007,Menezes,Moreira:2021uod}, which is introduced through the following assumption
\begin{equation}\label{A}
A^{\prime}=-\alpha W(\phi),    
\end{equation}
where $W(\phi)$ is the superpotential. 

We assume the function $f(Q)$ as being $f(Q)=Q+kQ^{n}$, which represents a good generalization for STEGR, where the parameters $k$ and $n$ represent the deviation from the usual theory. From field equations (\ref{e.q}) and (\ref{e.q2}), we obtain
\begin{eqnarray}
\phi^{\prime}&=&3\alpha[1+nC_n(\alpha W)^{2n-2}] W_\phi, \\
V(\phi)&=&\frac{9\alpha^2}{2} \Big[1+nC_n(\alpha W)^{2n-2}\Big]^2W_\phi^2-6[1+C_n(\alpha W)^{2n-2}](\alpha W)^{2},
\end{eqnarray}
where $C_n=12^{n-1}k(2n-1)$. One writes the energy density $\rho(y)=-e^{2A}\mathcal{L}_m$ in terms of the superpotential as
\begin{eqnarray}
\rho(y)=e^{2A(y)}(9\alpha^2[1+nC_n(\alpha W)^{2n-2}]^2 W_\phi^2-6[1+C_n(\alpha W)^{2n-2}](\alpha W)^{2}).
\end{eqnarray}

Thus, one can completely determine the thick brane system with a specific superpotential choice. In the sequel, we choose two kinds of superpotential to obtain the analytical expression for scalar field solution, warp factor, potential, and energy density. The first superpotential is the sine-Gordon one, for which we consider $n=1$. The second superpotential is linear one for $n=2$.

\subsection{Sine-Gordon superpotential}

Our first example is the sine-Gordon, whose superpotential is given by
\begin{equation}
W(\phi)=\beta^2\sin{\Big(\frac{\phi}{\beta}\Big)}.    
\end{equation}
Here, we make $n=1$. The scalar field solution and the potential read
\begin{eqnarray}\label{f1}
\phi(y)&=&\beta \arcsin\{\tanh[3\alpha y(1+k)]\}, \\
 V(\phi)&=&\frac{3}{2}(1+k)\alpha^2\beta^2\Big[3(1+k)\cos^2\Big(\frac{\phi}{\beta}\Big)-4\beta^2\sin^2\Big(\frac{\phi}{\beta}\Big)\Big].  
\end{eqnarray}
Substituting the solution (\ref{f1}) into (\ref{A}), we obtain the following expression for the warp factor
\begin{equation}
A(y)=\frac{\beta^2}{3(1+k)}\ln\Big\{\mathrm{ sech}[3\alpha(1+k)y]\Big\}.
\end{equation}
We can finally write the energy density as being
\begin{equation}
\rho(y)=3(1+k)\alpha^2\beta^2[\cosh(3\alpha y(1+k))]^{-\frac{2\beta^2}{3(1+k)}}\Big[3(1+k)-(3(1+k)+2\beta^2)\tanh^2(3\alpha y (1+k))\Big].    
\end{equation}

\begin{figure}[ht!]
\begin{center}
\begin{tabular}{ccc}
\includegraphics[height=4.5cm]{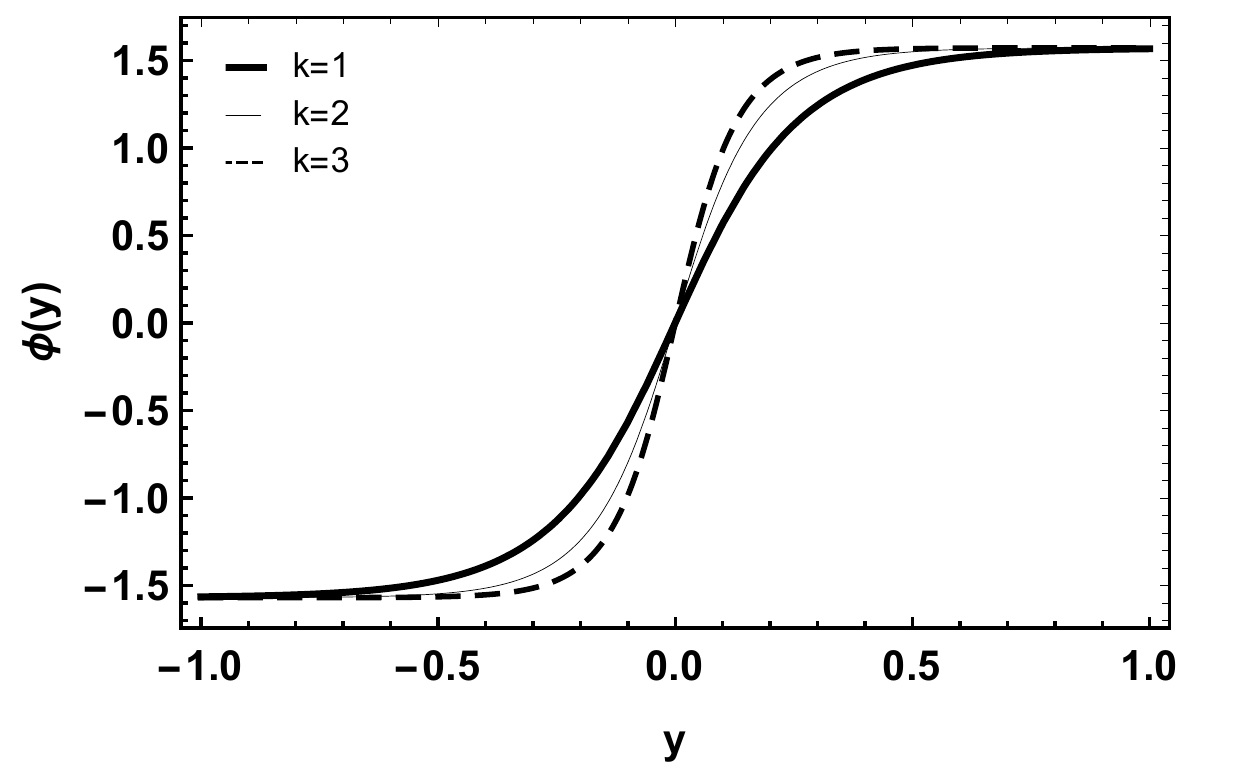} 
\includegraphics[height=4.5cm]{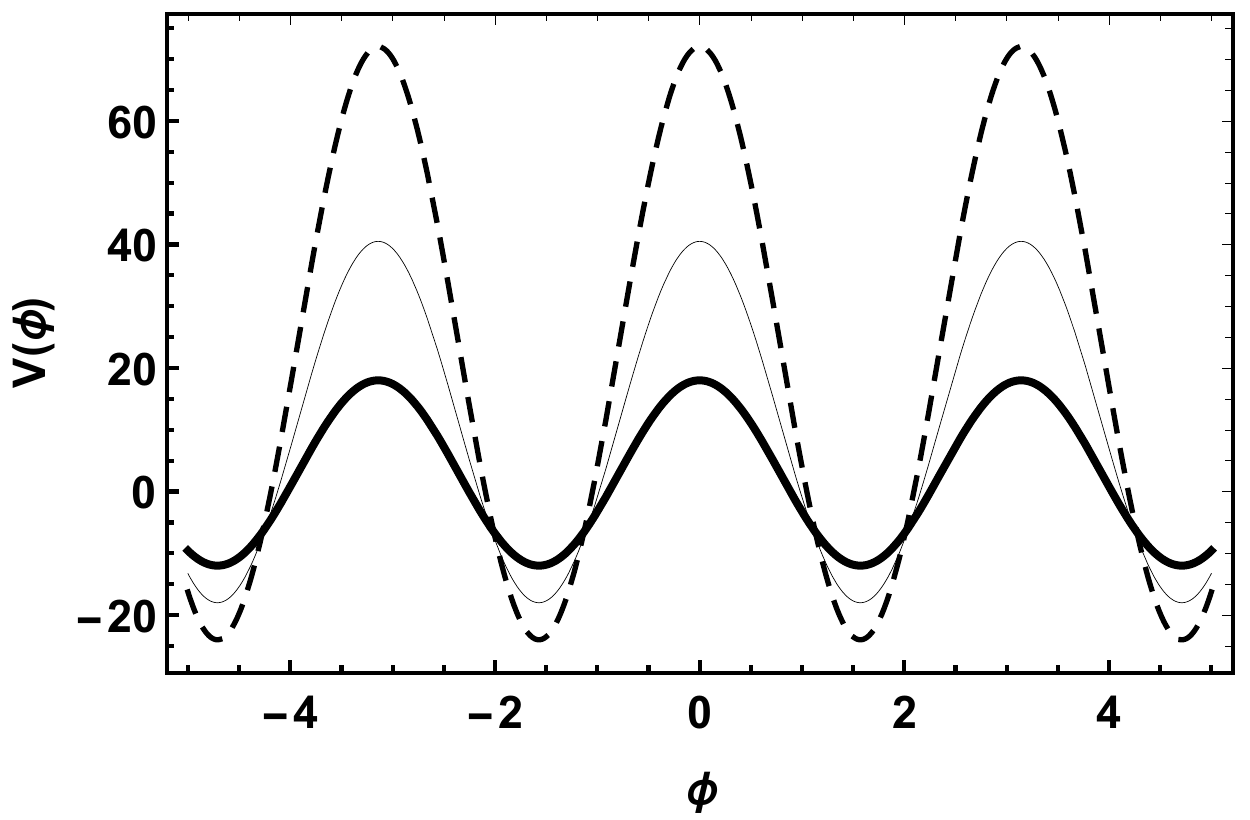}\\
(a) \hspace{7cm}(b)\\
\includegraphics[height=4.5cm]{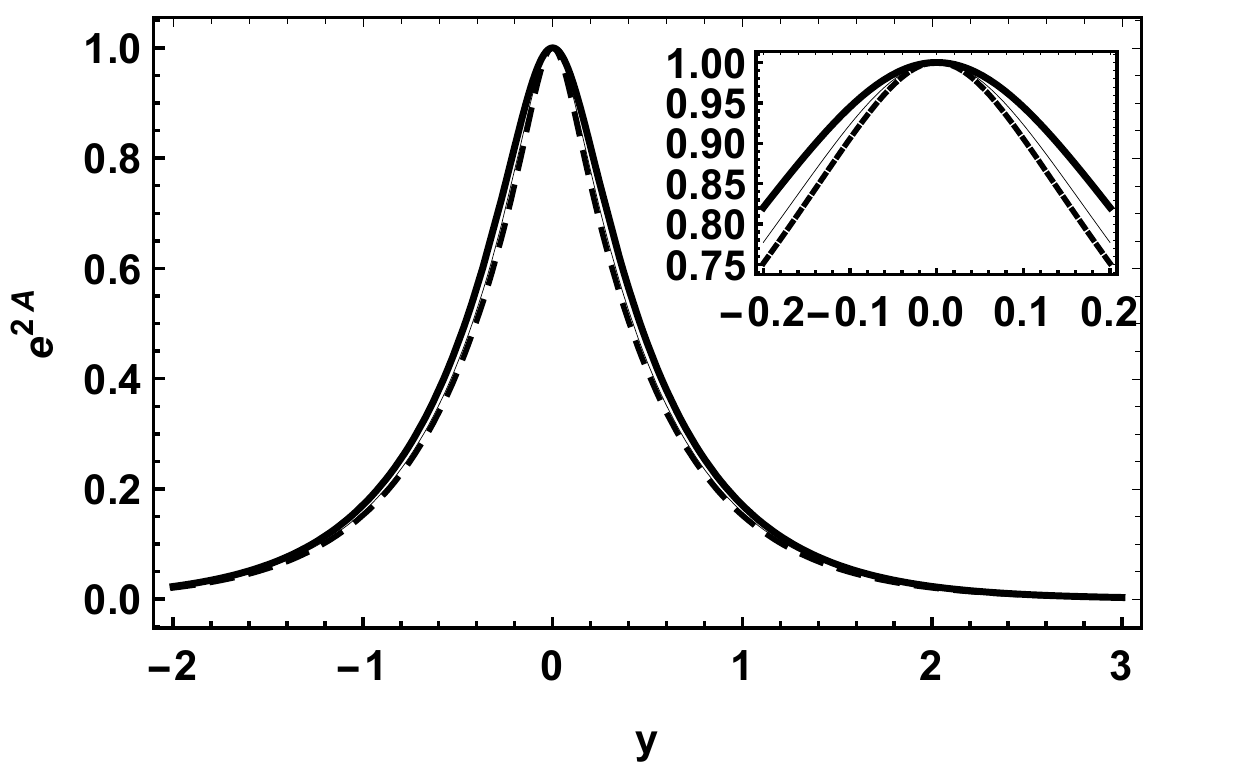} 
\includegraphics[height=4.5cm]{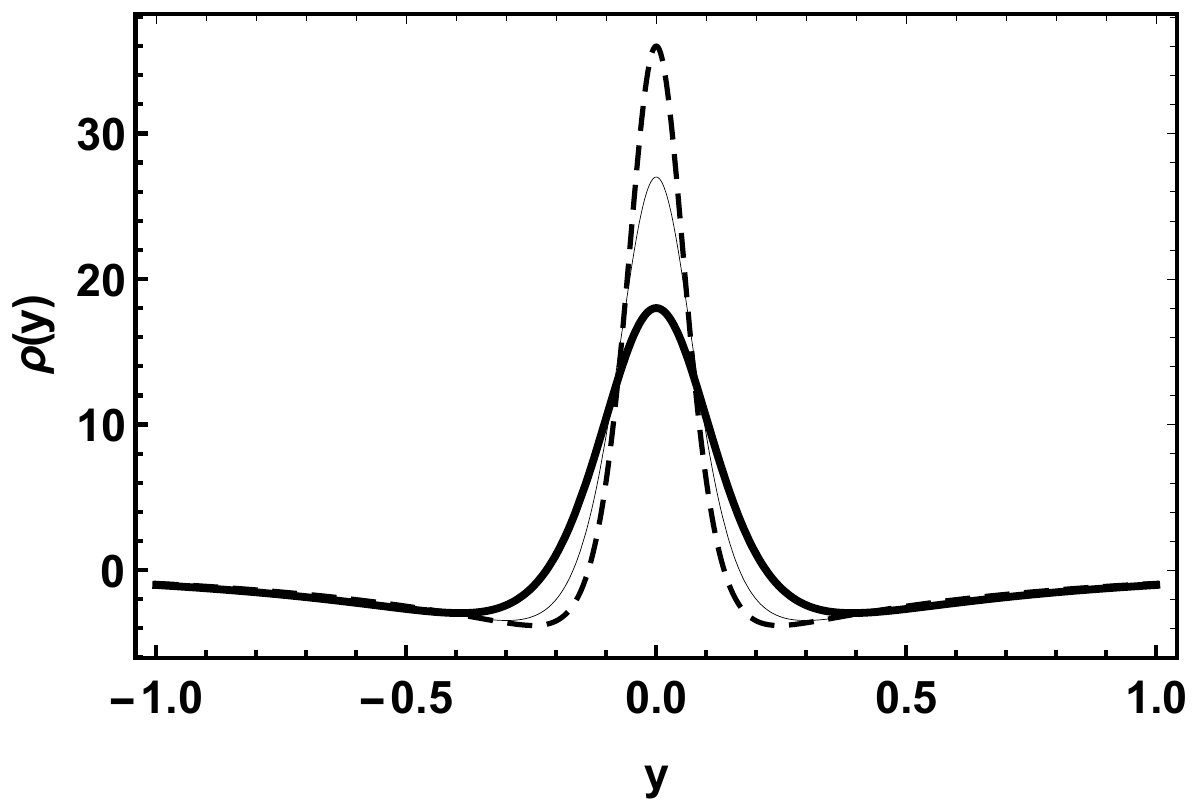}\\
(c) \hspace{7cm}(d)
\end{tabular}
\end{center}
\vspace{-0.5cm}
\caption{ For the sine-Gordon superpotential with $\alpha=\beta=1$. (a) Scalar field. (b) Potential. (c) Warp factor. (d) Energy density.
\label{fig1}}
\end{figure}

In Fig. \ref{fig1}, it is plotted the behavior for the scalar field solution $\phi$, potential $V(\phi)$, the warp factor $e^{2A}$ and the energy density $\rho$. As we can see, the scalar field exhibits a kink-like behavior, as expected. Besides, the potential has an oscillating behavior, and the energy density feels the variation of the parameter $k$, tending to become more localized as the parameter is increased.

\subsection{Linear superpotential}

We can consider linear example with the following superpotential
\begin{equation}
W(\phi)=\beta\phi.
\end{equation}
In this case, we take $n=2$ so that the scalar field for linear superpotential is
\begin{equation}
\phi(y)=\frac{\tan(18\sqrt{2k}\alpha^2\beta^2y)}{6\sqrt{2k}\alpha\beta},
\end{equation}
and the potential
\begin{equation}
V(\phi)=\frac{9}{2}\alpha^2\beta^2\Big[1+72k(\alpha\beta\phi)^2\Big]^2-6(\alpha\beta\phi)^2[1+36k(\alpha\beta\phi)^2].
\end{equation}
Besides, the warp factor is
\begin{equation}
A(y)=\frac{\ln[\cos(18\sqrt{2k}\alpha^2\beta^2y)]}{216k\alpha^2\beta^2},
\end{equation}
which brings us to an energy density of the form
\begin{eqnarray}
\rho(y)&=&\Big[\cos(18\sqrt{2k}\alpha^2\beta^2y)\Big]^{\frac{1}{108k\alpha^2\beta^2}}\Bigg\{9\alpha^2\beta^2(1+\tan^2(18\sqrt{2k}\alpha^2\beta^2y)^2\nonumber\\ &-&\frac{1}{24k}\tan^2(18\sqrt{2k}\alpha^2\beta^2y)\Big(2+\tan^2(18\sqrt{2k}\alpha^2\beta^2y)\Big)\Big]\Bigg\}.    
\end{eqnarray}

\begin{figure}[ht!]
\begin{center}
\begin{tabular}{ccc}
\includegraphics[height=4.5cm]{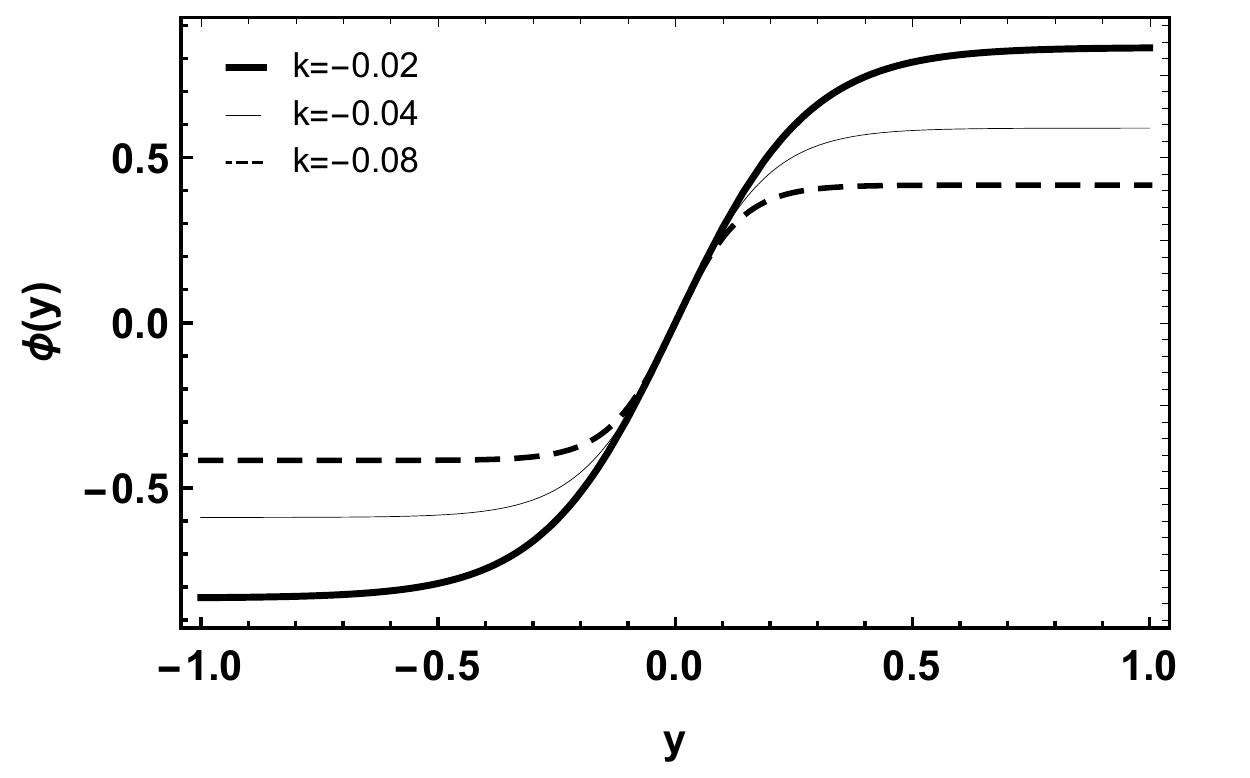} 
\includegraphics[height=4.5cm]{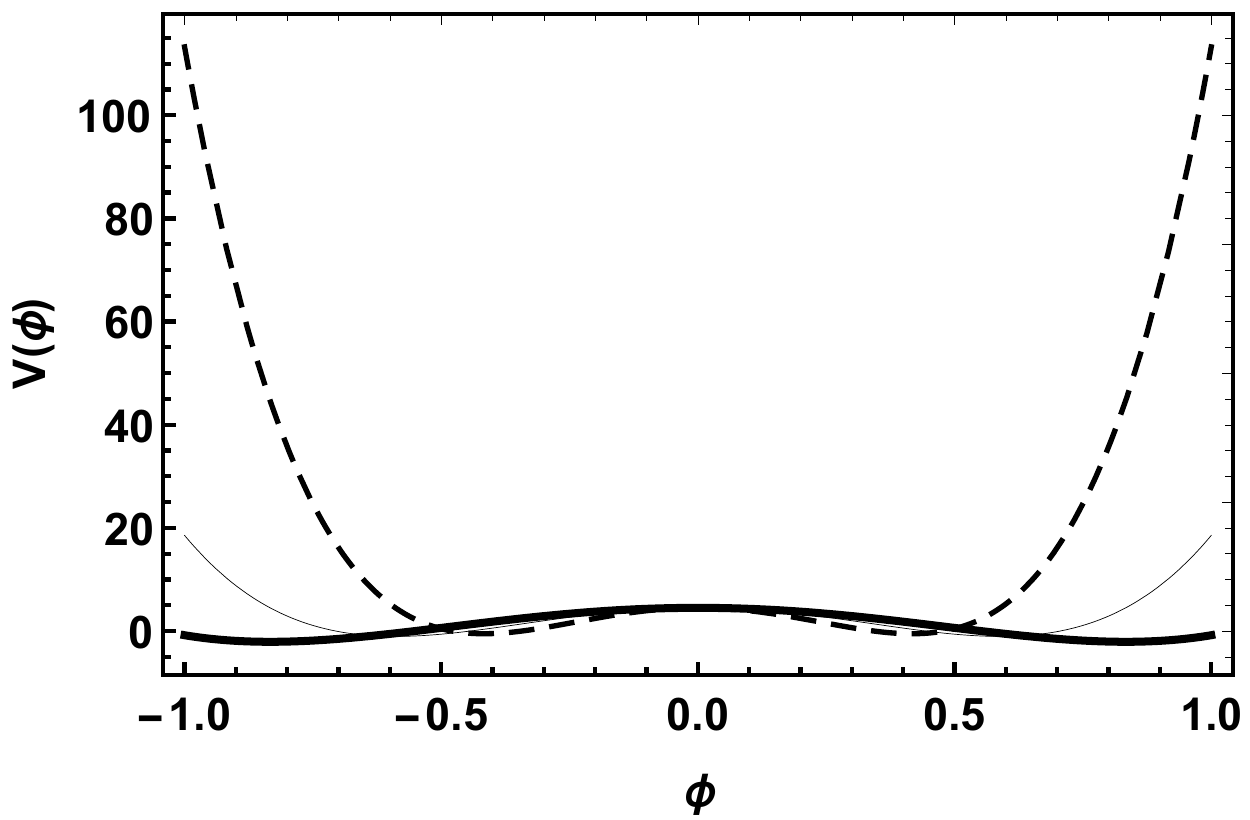}\\
(a) \hspace{7cm}(b)\\
\includegraphics[height=4.5cm]{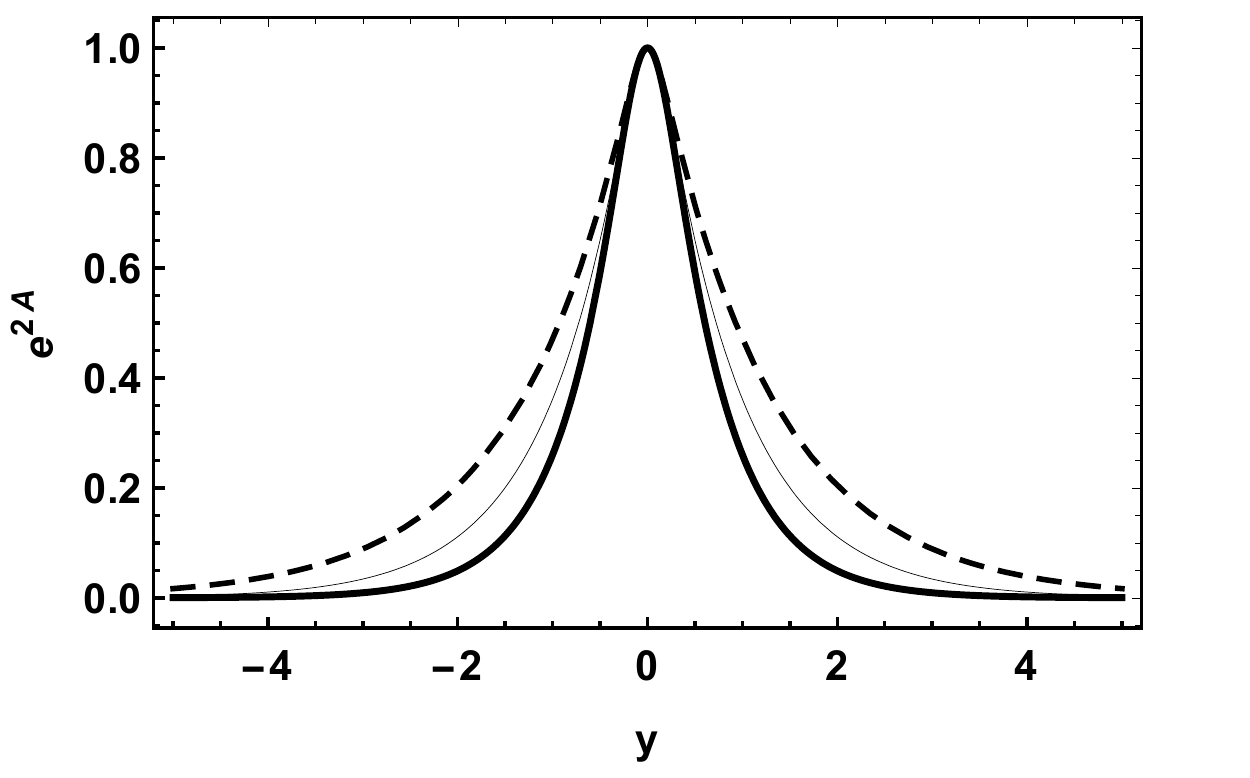} 
\includegraphics[height=4.5cm]{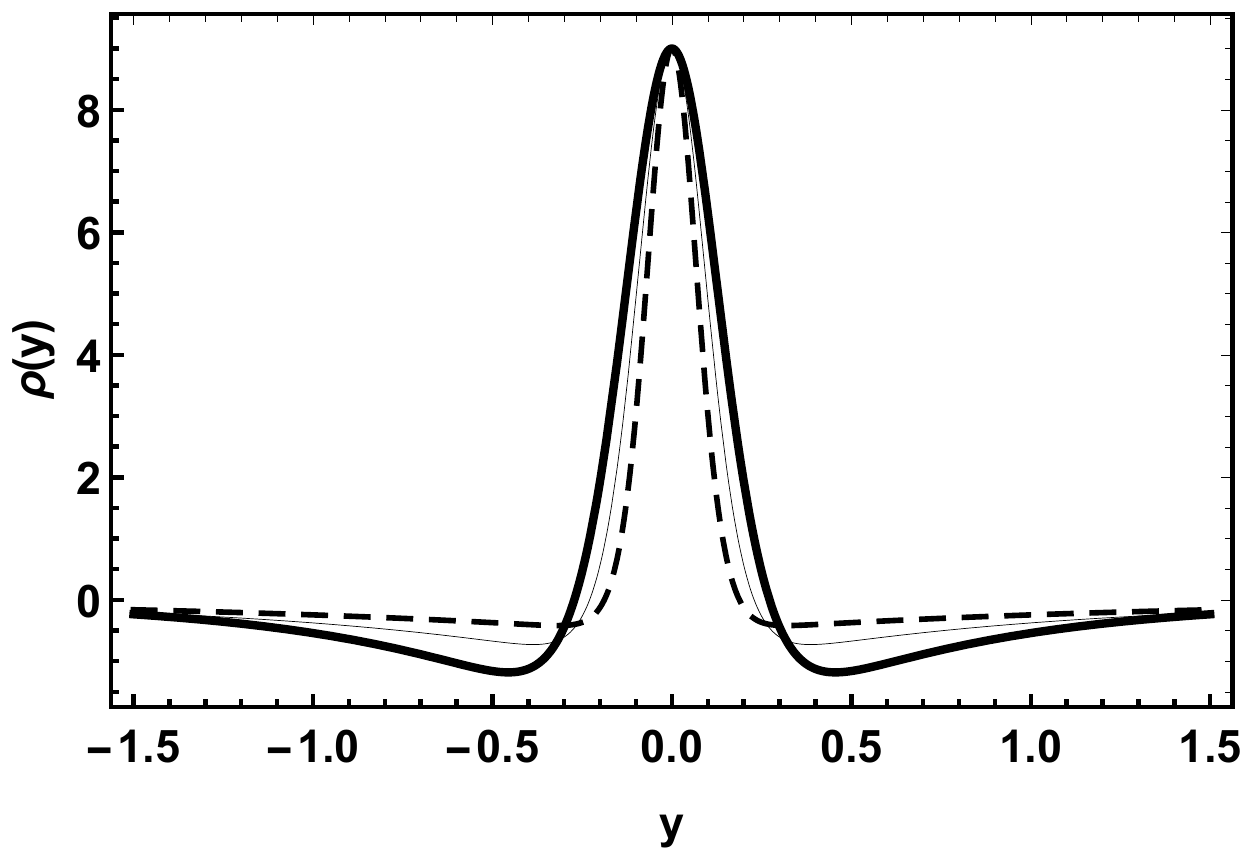}\\
(c) \hspace{7 cm}(d)
\end{tabular}
\end{center}
\vspace{-0.5cm}
\caption{For the linear superpotential with $\alpha=\beta=1$. (a) Scalar field. (b) Potential. (c) Warp factor. (d) Energy density.
\label{fig2}}
\end{figure}

From Fig. \ref{fig2}, we see that the scalar field solution for polynomial superpotential is also kink-like. Furthermore, the influence of the parameter $k$ on the solution of the scalar field $\phi$ and on the potential $V(\phi)$ is quite evident, directly affecting the energy density. It is interesting to note that the energy density becomes less localized as we choose a smaller value of $k$.

Assuming the thick brane constructed in $f(Q)$ gravity, the stability under small tensor pertubations and gravity localization was addressed in \cite{Fu:2021rgu}, while the localization of Dirac fermions with geometrical coupling was treated in \cite{Silva:2022pfd}.
Ahead we proceed to study the trapping of the 5D Elko field.
\section{Elko field localization}\label{s4}

Due to its properties, the Elko spinor could be a natural candidate for dark matter and deserves an investigation about its localization on braneworld.

\subsection{4D Elko spinor and dark matter}

Firstly, we introduce some features of Elko spinor in four-dimensional space-time. The Elko interacts weakly with matter fields and radiation as a mass dimension one spinor. Besides, there is observational evidence that suggests that dark matter is self-interacting. The Elko interaction is restricted to graviton and Higgs field. The term of self-interaction for a 4D Elko field can be accomplished by \cite{Ahluwalia:2010zn}
\begin{equation}\label{c1}
 g_{\lambda}[\overline{\lambda}(x)\lambda(x)]^2,  
\end{equation}
where $g_{\lambda}$ represents the dimensionless coupling constant and $\lambda(x)$ is the Elko quantum field 
\begin{equation}\label{elko}
\lambda(x)=\int\frac{d^4k}{(2\pi)^4}\frac{1}{\sqrt{2\omega_k}}\sum_\beta\Big[a_\beta(k)\lambda^A_\beta(k)  e^{-ikx}+a^{\dagger}_\beta(k)\lambda^A_\beta(k) e^{ikx}\Big].   
\end{equation}

The spin one-half eingenspinor $\lambda^{S/A}_\beta(k)$ must satisfy the relation $C\lambda^{S/A}_\beta(k)=\pm \lambda^{S/A}_\beta(k)$, where $C$ is the charge conjugation operator, $\lambda^S$ self conjugate (positive), $\lambda^A$ anti-self conjugate (negative), and $\beta=(\lbrace +,-\rbrace, \lbrace -,+\rbrace)$ represents the helicity. Also, this spinor and its dual satisfy the following orthonormality relations
\begin{eqnarray}
\overline{\lambda}^{S/A}_\beta(k)\lambda^{S/A}_\beta(k)&=&\pm 2m\delta_{\beta\beta},\nonumber\\
\overline{\lambda}^S_\beta(k)\lambda^A_\beta(k)=\overline{\lambda}^A_\beta(k)\lambda^S_\beta(k)&=&0.
\end{eqnarray}
Since the Elko field is fermionic one, the creation and annihilation operators must satisfy anti-commutation relations written below
\begin{eqnarray}
\lbrace a_\beta(k`),a_\beta(k)\rbrace=\lbrace a^{\dagger}_\beta(k`),a^{\dagger}_\beta(k)\rbrace&=&0,\nonumber\\
\lbrace a^{\dagger}_\beta(k`),a_\beta(k)\rbrace&=&\delta^{(3)}(k`-k)\delta_{\beta\beta}.
\end{eqnarray}

As we said before, the 4D Elko field can also be coupled to Higgs through the Yukawa-like term
\begin{equation}\label{c2}
 g_{\phi\lambda}[\phi^{\dagger}(x)\phi(x)][\overline{\lambda}(x)\lambda(x)],  
\end{equation}
where $g_{\phi\lambda}$ represents the dimensionless coupling constant between Elko and Higgs. For a more detailed review of the Elko field, we suggest to the reader the Refs.\cite{Ahluwalia:2008xi, Ahluwalia:2009rh, Ahluwalia:2010zn, Ahluwalia:2022ttu}.

\subsection{5D Elko field localization with Yukawa-like coupling}

Now, our interest is to verify whether the Elko field is trapped on the thick brane, which was built in the previous section. For this purpose, a nonminimal coupling between the Elko field and nonmetricity scalar is assumed. This is accomplished by a Yukawa-like interaction (similar to (\ref{c2})). Firstly, the conformal coordinate $dz=e^{-A}dy$ is introduced, so that the metric (\ref{metric}) is now written as follows
\begin{equation}\label{metric2}
 ds^2= e^{2A}(\eta_{\mu\nu}dx^{\mu}dx^{\nu}+dz^2).  
\end{equation}

In the five-dimensional spacetime, the action for the Elko field with a Yukawa-like coupling is given by \cite{Liu:2011nb, Jardim:2014xla}
\begin{equation}\label{ae}
S=\int d^5x \sqrt{-g}\Big[\frac{1}{4}(D_M\lambda D^M\overline{\lambda}+D_M\overline{\lambda} D^M\lambda)+ G(Q)\overline{\lambda}\lambda\Big],
\end{equation}
where $G(Q)$ is a suitable function of nonmetricity scalar employed to obtain a normalizable mode for the Elko field. From action (\ref{ae}), we observe that the 5D Elko spinor has mass dimension 3/2 unlike the 4D Elko, which has mass dimension one. In addition, we define the covariant derivative in symmetric teleparallel gravity as $D_M=\partial_M+\Omega_M$, where $\Omega_M$ is the torsion-free spin connection defined by
\begin{equation}\label{con}
 \Omega_M=\frac{1}{4} \omega_M^{ab} \gamma_{a}\gamma_{b}, 
\end{equation}
where $\gamma_{a}$ represents the Dirac matrix in flat spacetime.
To obtain $\omega_M^{ab}$, we use the Cartan equation $d\theta^a+\omega_b^a\theta^b=0$ with the torsion-free condition. Recall that $\theta^a=h^a_M d^Mx$ and $\omega_b^a=\omega^a_{bM} d^Mx$, where $h^a_M$ is the vielbein that satisfy the relation $g_{MN}=\eta_{ab}h^a_M h^b_N$. We should point out that the spin connection is preserved since we deal with vanishing torsion. Thus, for the metric (\ref{metric}), we have the only non vanishing connection $\Omega_\mu=\frac{1}{2} \dot{A} \gamma_{\mu}\gamma_{4}$. Here, the dot stands for the derivative with respect to the conformal coordinate, i.e., $d/dz$.

The corresponding equation of motion obtained from the action (\ref{ae}) is 
\begin{equation}
D_M(\sqrt{-g}D^M\lambda) - 2\sqrt{-g} G(Q)\lambda=0. 
\end{equation}
It can be shown that using the metric (\ref{metric2}) and spin connection (\ref{con}), the above equation becomes 
\begin{equation}
 \square\lambda-\dot{A}\gamma^4\gamma^\mu\partial_\mu-\dot{A}^2\lambda+\partial_4^2\lambda+3\dot{A}\partial_4\lambda-2e^{2A} G\lambda=0,
\end{equation}
where $\square=\eta^{\mu\nu}\partial_\mu\partial_\nu$. 

Similar to the solution (\ref{elko}), it is possible to adopt a Kaluza-Klein decomposition for the Elko field as
\begin{equation}
 \lambda_{\pm}(x^\mu,z)=\sum_{n,\beta}\chi_n(z)[\lambda^A_{n,\beta}(x)+\lambda^S_{n,\beta}(x)],   
\end{equation}
implying the following equation for $\chi(z)$
\begin{equation}\label{eqE}
\ddot{\chi}+3\dot{A}\dot{\chi}-\left(\dot{A}^2+im\dot{A}+ 2e^{2A} G\right)\chi=-m^2\chi.    
\end{equation}
To obtain (\ref{eqE}), we assume that the 4D Elko spinor satisfies the following relations
\begin{eqnarray}
\gamma^\mu\partial_\mu\lambda^A_{n,\pm}(x)=\mp i\lambda^A_{n,\mp}(x), \ \gamma^\mu\partial_\mu\lambda^S_{n,\pm}(x)=\pm i\lambda^S_{n,\pm}(x),\\
\gamma^4\lambda^A_{n,\pm}(x)=\pm\lambda^S_{n,\mp}(x), \ \gamma^4\lambda^S_{n,\pm}(x)=\mp\lambda^A_{n,\mp}(x),\\
\square\lambda^A_{n,\pm}(x)=m^2\lambda^A_{n,\pm}(x), \
\square\lambda^S_{n,\pm}(x)=m^2\lambda^S_{n,\pm}(x),
\end{eqnarray}
where $m$ represents the 4D Elko spinor mass.

The next step is to transform the equation (\ref{eqE}) into the Schr\"{o}dinger-like form. For this purpose, we must make the following change $\chi(z)=e^{-\frac{3}{2}A(z)}\psi(z)$ that implies in
\begin{equation}\label{33333}
 -\ddot{\psi}+V\psi=m^2\psi,  
\end{equation}
where the effective potential reads
\begin{equation}\label{pee}
 V=\frac{3}{2}\ddot{A}+\frac{13}{4} \dot{A}^2+ i m \dot{A}+ 2e^{2A}G.
\end{equation}

It is convenient to take the function $G$ as $G(Q)=c Q$ to investigate zero-mode localization ($m=0$). Then the potential (\ref{pee}) is rewritten as
\begin{equation}\label{pe}
V=\frac{3}{2}\ddot{A}+\left(\frac{13}{4} +24 c \right)\dot{A}^2+im \dot{A}.    
\end{equation}
Here, $c$ is a parameter to be chosen in a way that allows the localization of the Elko field zero-mode on the brane.

\subsubsection{Zero-mode}

To obtain a factorized potential, we must have $c=-\frac{1}{24}$, so that (\ref{pe}) for $m=0$ reduces to 
\begin{equation}\label{vef}
V=\frac{3}{2}\ddot{A}+\frac{9}{4}\dot{A}^2.    
\end{equation}

It is worth mentioning that the above potential is the same as the scalar field. Thus, the zero-mode takes the simple form
\begin{equation}
 \psi_0(z)=N_0 e^{\frac{3}{2}A(z)}.  
\end{equation}

\begin{figure}[ht!]
\begin{center}
\begin{tabular}{ccc}
\includegraphics[height=4.5cm]{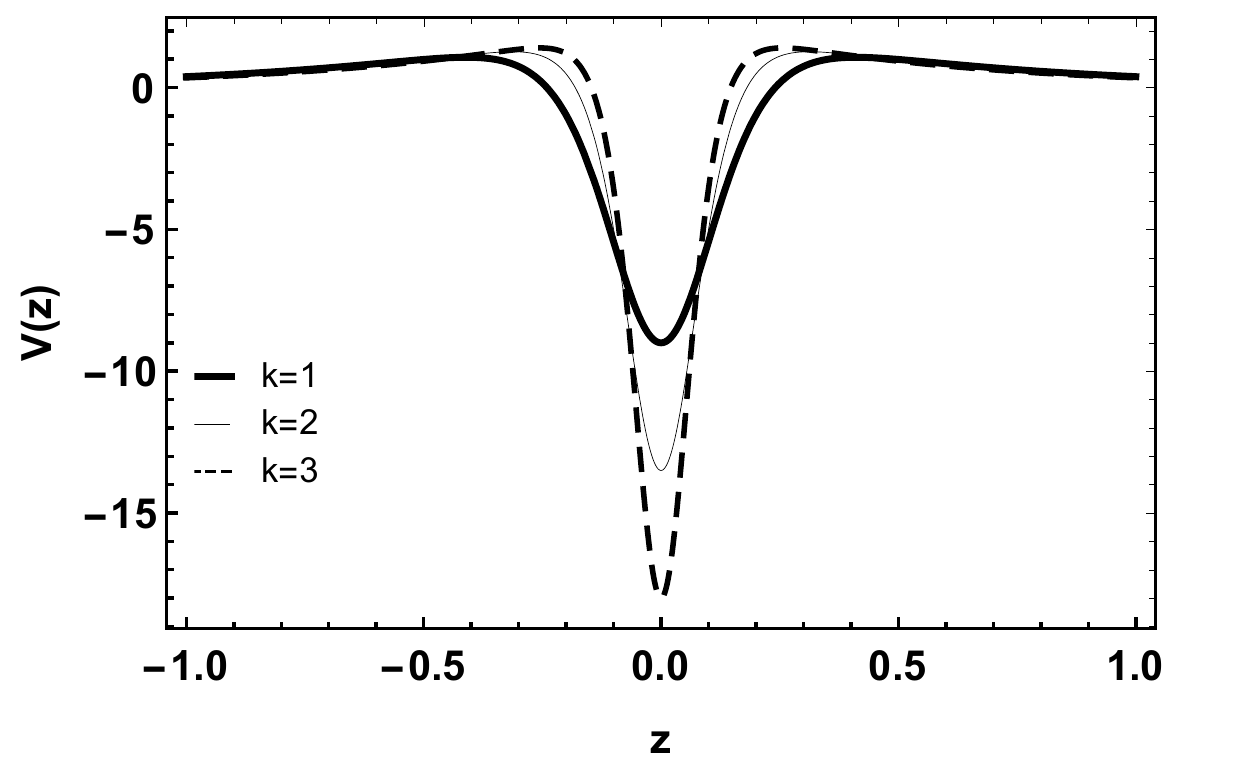} 
\includegraphics[height=4.5cm]{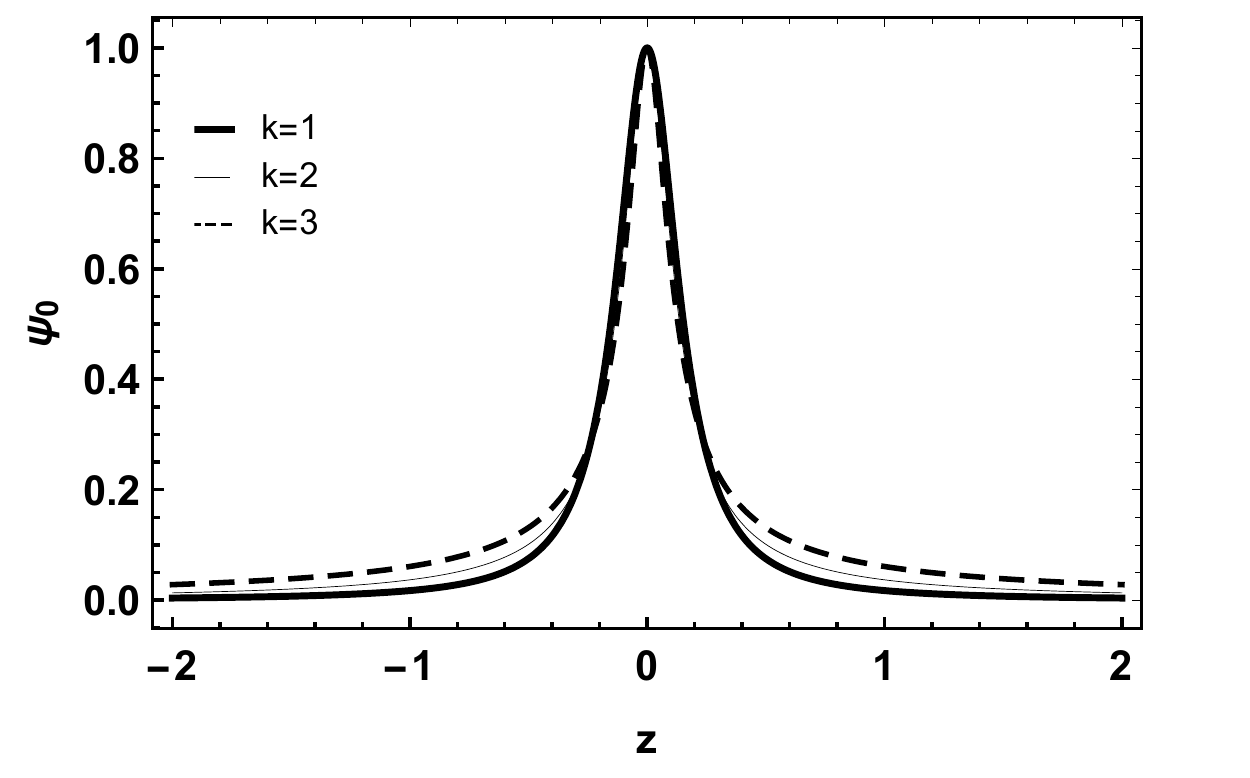}\\
(a) \hspace{7cm}(b)\\
\includegraphics[height=4.5cm]{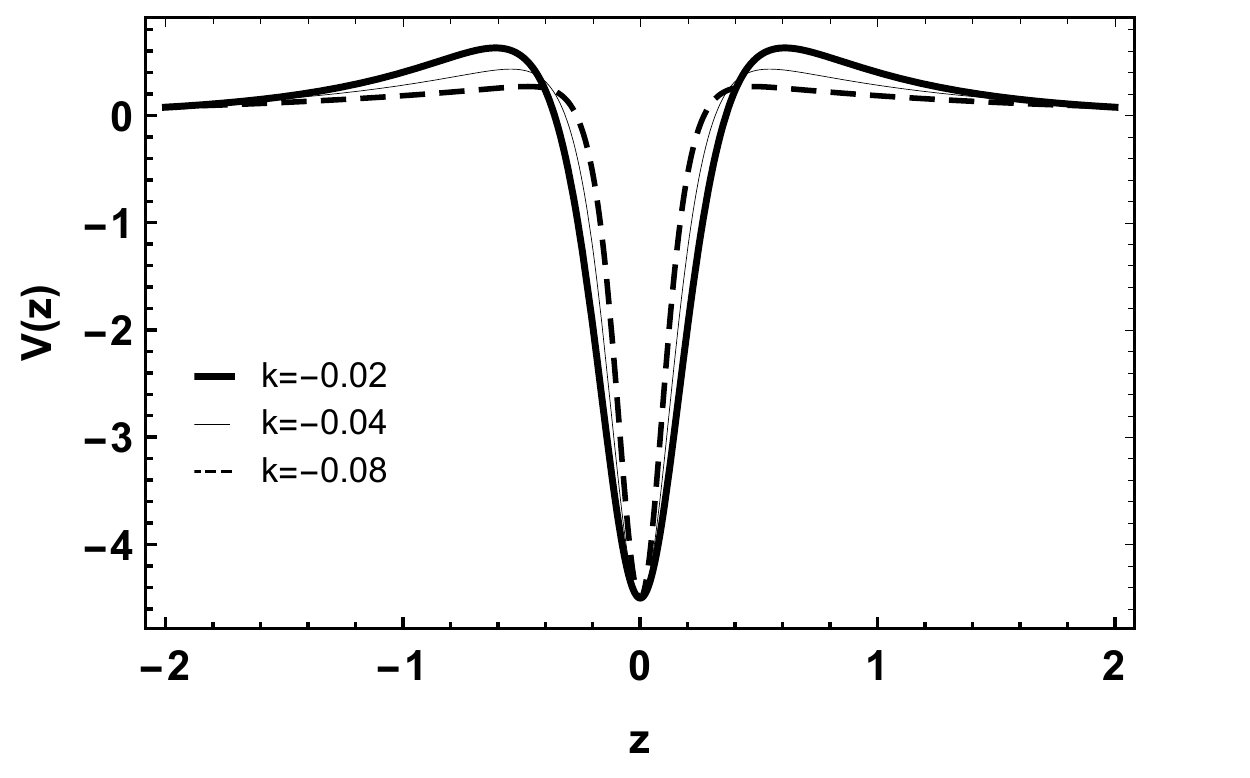} 
\includegraphics[height=4.5cm]{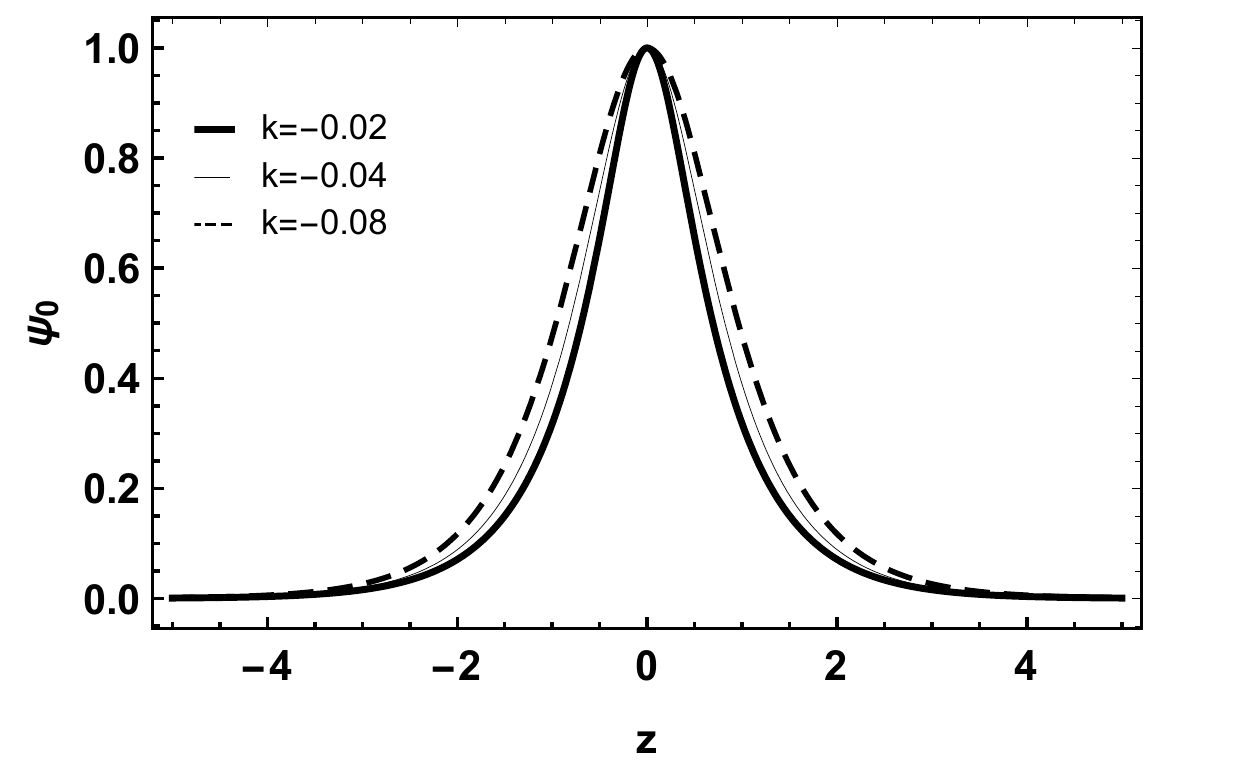}\\
(c) \hspace{7 cm}(d)
\end{tabular}
\end{center}
\vspace{-0.5cm}
\caption{ The shape of the potential and of the zero-mode with $\alpha=\beta=1$. (a) and (b) sine-Gordon superpotential. (c) and (d) linear superpotential.
\label{fig3}}
\end{figure}

In Fig. \ref{fig3}, we plot the behavior of the effective potential and the zero-mode for the sine-Gordon type superpotential (Fig.\ref{fig3}a and b), and for the linear superpotential (Fig.\ref{fig3}c and d).  In both cases, when we increase the value of the $k$ parameter, the potential well intensifies, making the zero-modes more localized.

\subsubsection{Massive modes}

It would be considerably hard to study the massive spectrum since the effective potential (\ref{pe}) is complex. Inspired by 6D Elko on string-like brane \cite{Dantas:2015mfi}, we add the exotic term $-\frac{ime^{-A}}{4\sqrt{3}}\sqrt{Q}$ to the function $G(Q)$ in order to remove the imaginary part of the effective potential. Then, for massive modes, the effective potential also assumes the form (\ref{vef}). As this effective potential is factorized, we can undoubtedly say that the massive spectrum has no tachyonic modes. 

\begin{figure}[ht!]
\begin{center}
\begin{tabular}{ccc}
\includegraphics[height=4.5cm]{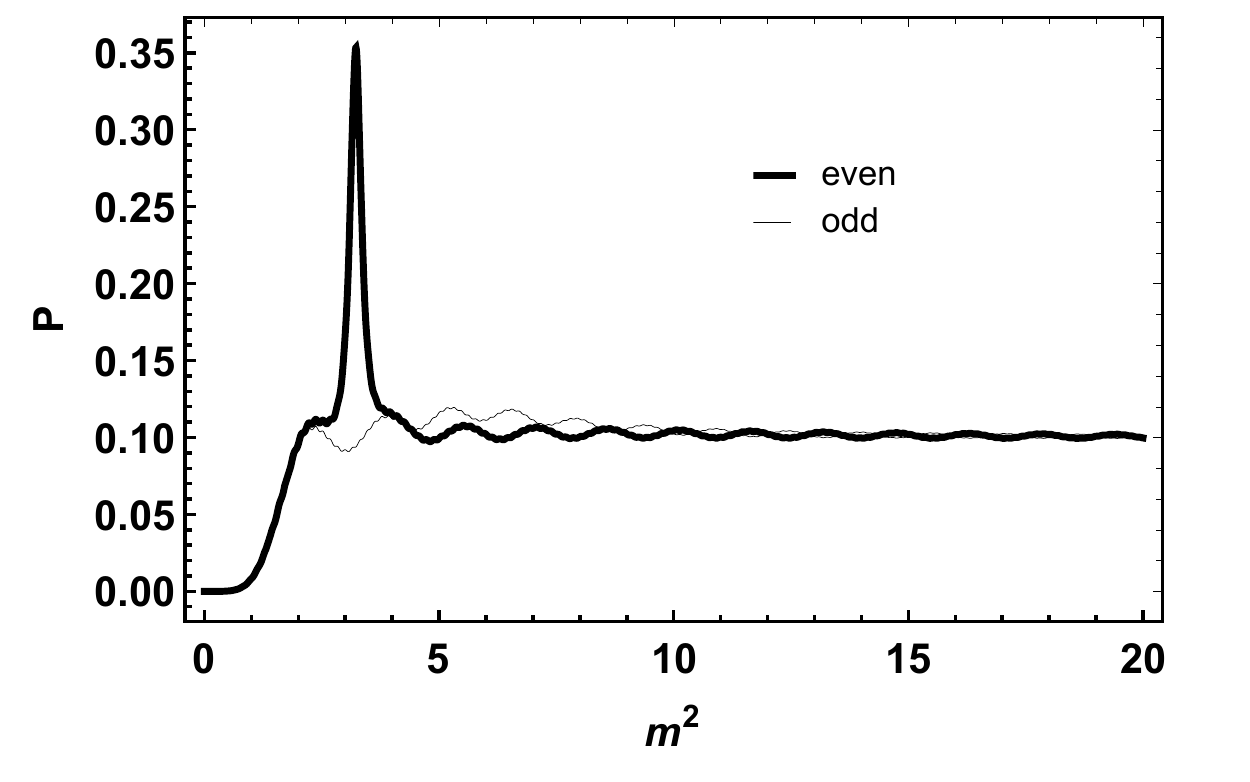} \\
(a)\\
\includegraphics[height=4.5cm]{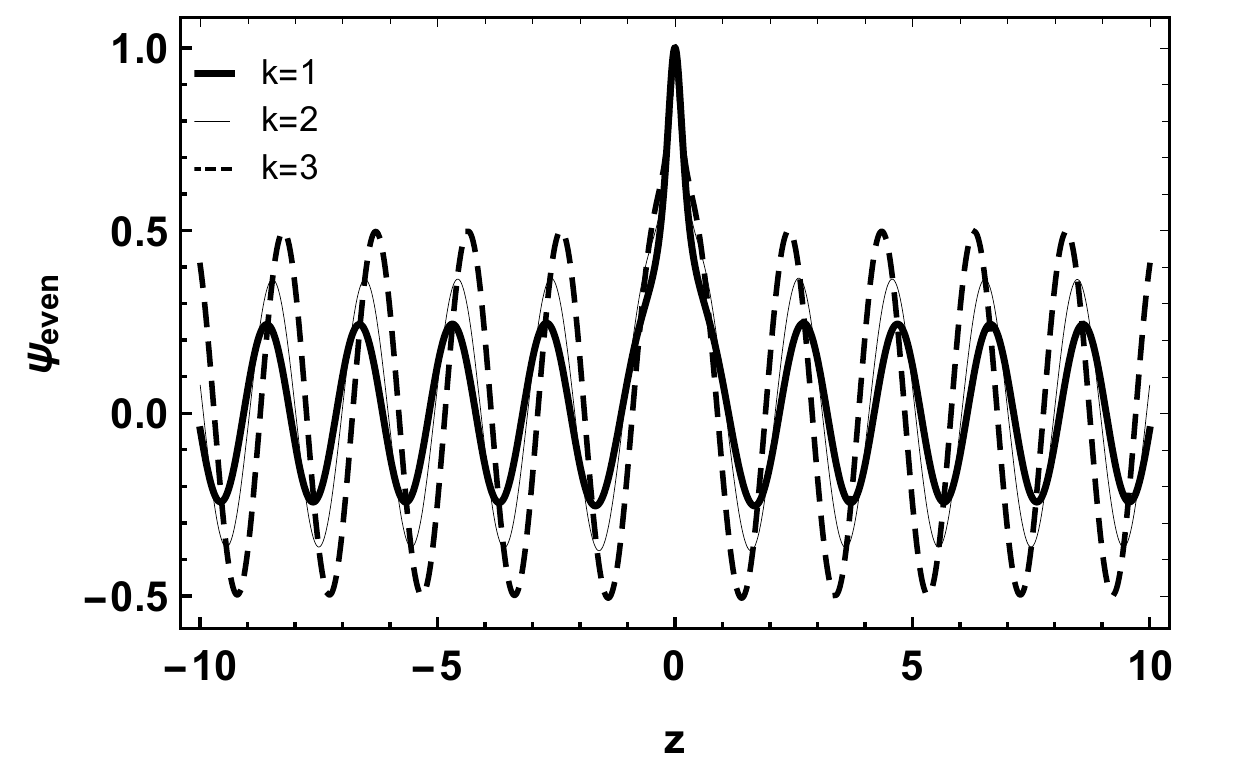} 
\includegraphics[height=4.5cm]{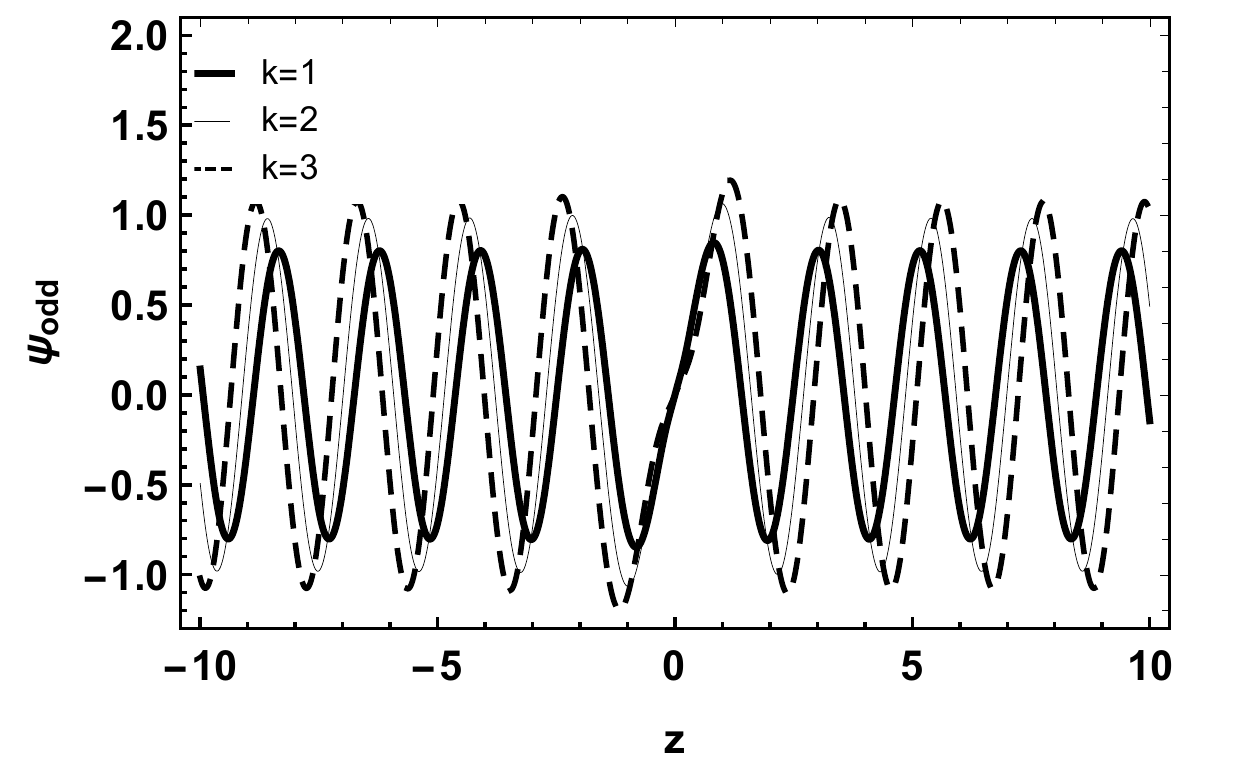}\\
(b) \hspace{7 cm}(c)
\end{tabular}
\end{center}
\vspace{-0.5cm}
\caption{The shape of the relative probability and of the massive modes with $\alpha=\beta=1$ for sine-Gordon superpotential. (a) $k=1$. (b) $m^2=3.235$.  (c)  $m^2=2.293$.
\label{fig5}}
\end{figure}

\begin{figure}[ht!]
\begin{center}
\begin{tabular}{ccc}
\includegraphics[height=4.5cm]{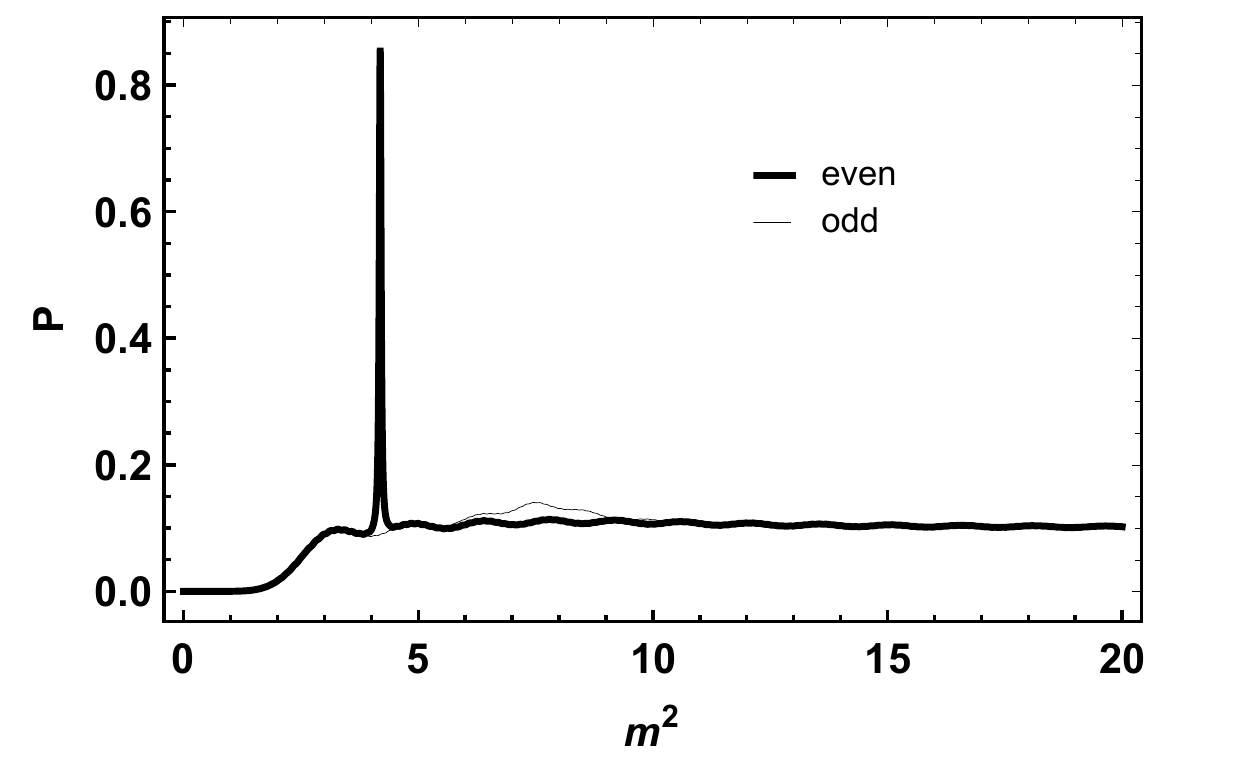} \\
(a)\\
\includegraphics[height=4.5cm]{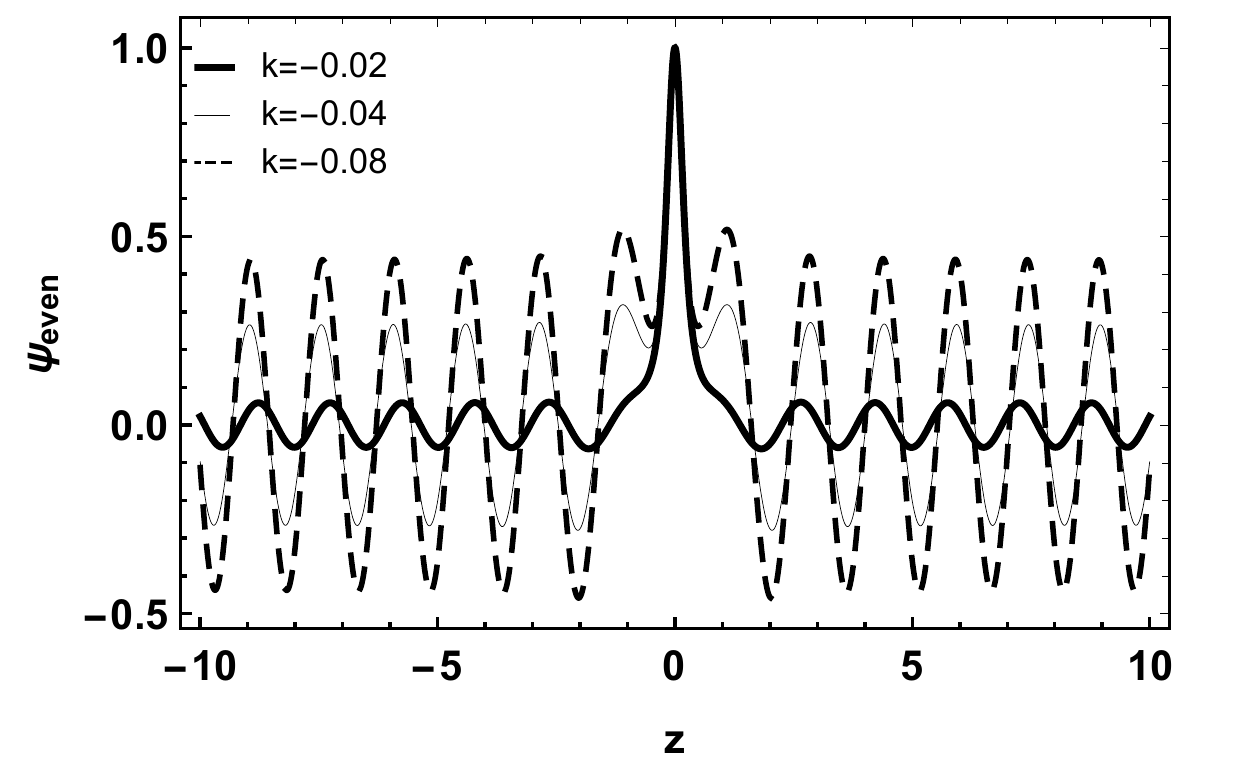} 
\includegraphics[height=4.5cm]{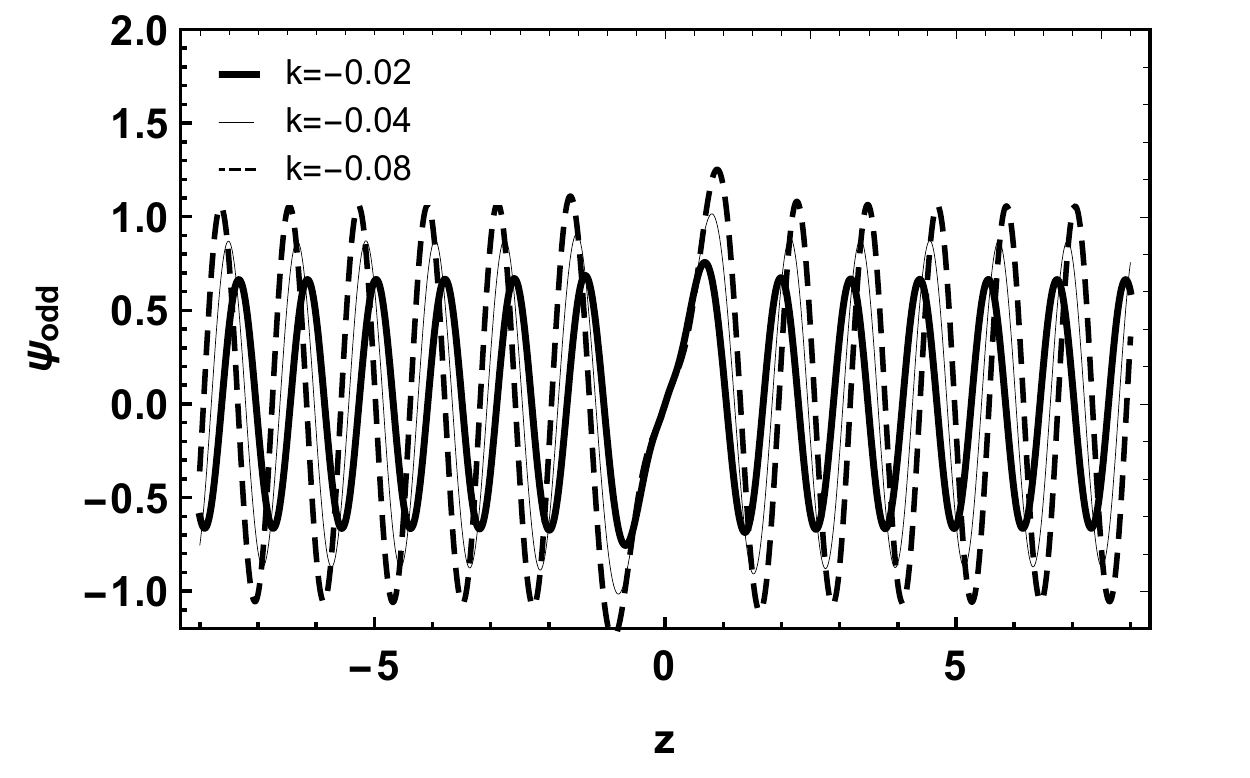}\\
(b) \hspace{7 cm}(c)
\end{tabular}
\end{center}
\vspace{-0.5cm}
\caption{The shape of the relative probability and of the massive modes with $\alpha=\beta=1$ for linear superpotential. (a) $k=-0.02$. (b) $m^2=4.186$.  (c)  $m^2=3.344$.
\label{fig6}}
\end{figure}

Notice that Eq.(\ref{33333}) can only be solved numerically, even eliminating the imaginary part. To do this, we use the interpolation method and assume the boundary conditions: $\psi_{even}(0)=1, \dot{\psi}_{even}(0)=0$ for even modes and $\psi_{odd}(0)=0, \dot{\psi}_{odd}(0)=1$ for odd modes \cite{Liu:2009ve}. We choose the boundary conditions on account of the behavior of the effective potentials $V(z)$ (Fig.\ref{fig3} a and c) which are of even functions, ensuring that the solutions will be wave functions even $\psi_{even}$ or odd $\psi_{odd}$. As expected, we see from (Fig.\ref{fig5}b and c) and (Fig.\ref{fig6}b and c) that the massive eigenfunctions have different behaviors near the origin but all exhibit periodic behavior far from the origin.

Furthermore, we can analyze the resonant modes, which are the massive modes that exhibit a respectively large amplitude near the brane \cite{Moreira:2021wkj,Ahluwalia:2022ttu}. To identify the resonance modes, it is necessary to calculate the relative probability $P(m)$ of finding a particle with a respective mass $m$ in a narrow band $2z_b$ \cite{Liu:2009ve,Liu2009a,Tan:2020sys}
\begin{eqnarray}
P(m)=\frac{\int_{-z_b}^{z_b} |\psi(z)|^2 dz}{\int_{-z_{max}}^{z_{max}} |\psi(z)|^2 dz}.
\end{eqnarray}
Here, $z_{max}$ represents the limit of the domain. It is interesting to note that the choice of parameter $z_b$ does not change the positions of the resonance peaks. However, with the smaller value of the $z_b$ parameter, it becomes easier to identify the resonance peaks.

At this point, we must highlight that the behavior of the potential shown in Fig. \ref{fig3} would commonly not support resonances. However, when analyzing the relative probabilities we find resonant modes for the even solutions (Fig.\ref{fig5}b and Fig.\ref{fig6}b). Furthermore, the relative probabilities give us a complete view of the behavior of the massive modes. As we can observe, for the Sino-Gordon superpotential, the first peak of the relative probability represents a massive mode that presents greater amplitude close to the brane (Fig.\ref{fig5}a). The same goes for the linear superpotential (Fig.\ref{fig6}a).

\section{Final remarks and Perspectives}\label{s5}

In this paper, we have shown that the Elko field can be confined on a thick brane in $f(Q)$ modified symmetric teleparallel gravity through a Yukawa-like interaction between the field and the nonmetricity scalar. Such a coupling allows us to obtain a normalizable zero-mode, besides providing a real-valued effective potential that enables us to study massive and resonates modes through the Schr\"{o}dinger approach. 

Employing the first-order formalism, we have built the brane system where it has been used the function $f(Q)=Q+kQ^n$. For $n=1$, we have considered a Sine-Gordon superpotential, whereas a linear one for $n=2$. In both cases, only the Elko zero-mode was shown to be confined on brane. We have plotted the behavior of effective potential and zero-mode, offering the same results as the scalar field. We have also plotted the massive and resonates modes for $n=1$ and $n=2$. Through Fig. \ref{fig5}(a) and Fig. \ref{fig6}(a), it is possible to see more clearly the existence of resonant modes, but these modes exist only in even solutions.

Our results represent a generalization of works where only the massless modes were analyzed. Besides, it is worth emphasizing that all previous works on the Elko field localization were carried out in the context of general relativity. For the first time, the influence of nonmetricity on the trapping of Elko spinor on braneworld is investigated.

For future works, we could study the localization of the Elko field by considering a dilaton-like geometrical coupling in which the function $G(Q)$ would be directly introduced in the Elko kinetic term. With such a coupling, an exotic term might not be necessary to obtain real-valued massive and resonates modes. Furthermore, this investigation could be extended to higher codimensions.

\section*{Acknowledgments}
\hspace{0.5cm} The authors thank the Funda\c{c}\~{a}o Cearense de Apoio ao Desenvolvimento Cient\'{i}fico e Tecnol\'{o}gico (FUNCAP), the Coordena\c{c}\~{a}o de Aperfei\c{c}oamento de Pessoal de N\'{i}vel Superior (CAPES), and the Conselho Nacional de Desenvolvimento Cient\'{i}fico e Tecnol\'{o}gico (CNPq), Grants no. 200879/2022-7 (RVM) and no. 309553/2021-0 (CASA) for financial support. R. V. Maluf acknowledges the Departament de F\'{i}sica Te\`{o}rica de la Universitat de Val\`{e}ncia for the kind hospitality. The authors also thank the anonymous referees for their valuable comments and suggestions.

\end{document}